\documentclass{article}

\usepackage{microtype}
\usepackage{graphicx}
\usepackage{etoolbox}
\usepackage{placeins}
\usepackage{subfigure}
\usepackage{booktabs}
\usepackage{amsmath}
\usepackage{xcolor}
\usepackage{xspace}
\newcommand{\sysname}{Penelope\xspace}
\newcommand{\mytitle}[1]{\vspace{2pt}\noindent\textbf{#1.}\xspace}

\usepackage{hyperref}

\usepackage[accepted]{mlsys2025}
\makeatletter
\renewcommand{\Notice@String}{Preprint.}
\makeatother
\hypersetup{pdfsubject={}}

\mlsystitlerunning{Rethinking KV Cache Eviction for Agentic LLM Serving}

\begin{document}

\twocolumn[{
\mlsystitle{Rethinking KV Cache Eviction for Agentic LLM Serving}

\begin{mlsysauthorlist}
\mlsysauthor{Jiakun Fan}{vt}
\mlsysauthor{Han Meng}{ucm}
\mlsysauthor{Yuxuan Xia}{nyu}
\mlsysauthor{Hao Mark Chen}{icl}
\mlsysauthor{Dong Li}{ucm}
\mlsysauthor{Dimitrios S. Nikolopoulos}{vt}
\end{mlsysauthorlist}

\mlsysaffiliation{vt}{Virginia Polytechnic Institute and State University}
\mlsysaffiliation{ucm}{University of California, Merced}
\mlsysaffiliation{nyu}{New York University}
\mlsysaffiliation{icl}{Imperial College London}
\mlsyscorrespondingauthor{Jiakun Fan}{jiakunfan@vt.edu}
\mlsyscorrespondingauthor{Dimitrios S. Nikolopoulos}{dsn@vt.edu}

\mlsyskeywords{KV cache eviction, LLM serving, memory management, agent workloads}

\vskip 0.3in

\begin{abstract}
Key-value (KV) caches impose substantial memory pressure on large
language model serving systems, requiring cache eviction when capacity
is insufficient. Existing mechanisms typically evict KV from the ends
of cached sequences to preserve prefix sharing across requests.
However, agent workloads repeatedly reuse long
histories across successive turns. Recomputing the KV cache for tokens
near the end of a sequence is substantially more expensive than for
earlier tokens because their attention computation involves longer
histories. To address this problem,
we propose preserving shared prefixes and evicting the KV cache from the
beginning of the remaining history while retaining the cached suffix.
We build \sysname
around this strategy by co-designing the eviction policy, KV
recomputation, and unified GPU--CPU cache management. Returning requests
reuse the KV cache retained in GPU or CPU memory and recompute only missing
segments while preserving reuse of the cached suffix. We implement
\sysname in SGLang and evaluate it on three Qwen3 model sizes using
real agent traces.
Compared with SGLang, \sysname reduces mean time to first token by up
to 60\% and job completion time by up to 37\% with a GPU-only cache,
and by up to 76\% and 60\% with a two-tier GPU--CPU cache.
\end{abstract}
}]

\printAffiliationsAndNotice{}

\section{Introduction}
\label{sec:intro}

Large language models generate tokens
autoregressively~\cite{vaswani2017attention}, with each new
token attending to the preceding context. To avoid recomputing the keys
and values of previous tokens at every decoding step, serving systems
store them in a key-value (KV) cache. The cache grows with sequence
length, and serving many requests concurrently places substantial
pressure on GPU memory. Beyond reuse during decoding, systems such as
vLLM~\cite{kwon2023vllm} and SGLang~\cite{zheng2024sglang} retain KV cache
after a request completes, allowing subsequent requests with matching
prefixes to reuse previously computed results. Retaining these histories
saves computation but also consumes limited cache capacity, making
eviction necessary as new requests compete for space.

Cache eviction involves two decisions: which cached histories to reclaim
space from, and which portions of those histories to remove. Mainstream
serving engines, including vLLM~\cite{kwon2023vllm} and
SGLang~\cite{zheng2024sglang}, use least recently used (LRU) policies to
guide cache eviction. Concurrent work further shows that more sophisticated
replacement policies often provide limited gains over LRU for prefix
caching~\cite{liu2026fancyeviction}. While LRU prioritizes cached content
by recency, it does not by itself specify which portion of a cached
sequence to remove. Existing engines typically evict KV from the ends
of cached sequences: vLLM releases a request's blocks starting from its
tail~\cite{vllm2026apc}, and SGLang removes leaf nodes of its radix
tree~\cite{zheng2024sglang}. This choice follows the structure of prefix reuse:
a cached token can be reused only when the request matches the prefix
through that position. Requests that diverge later can still share the
earlier prefix. Evicting from the end therefore preserves a contiguous
prefix that can be reused by a broader set of requests.

However, evicting from the end removes the tokens that are most
expensive to recompute. When evicted history is accessed again, its
missing KV cache must be recomputed. Under causal attention, each token
attends to all preceding tokens, so tokens near the end of a sequence
require more attention computation to recompute than earlier tokens.
Prior work exploits this asymmetry.
Pensieve~\cite{yu2025pensieve} evicts from the beginning of a
conversation's history in multi-turn chat;
AsymCache~\cite{shi2026asymcache} weighs hit rate against
position-dependent recomputation cost when evicting blocks; and a
trace study of prefix caching~\cite{liu2026fancyeviction} recommends
partially evicting cached nodes from their beginning.

\begin{figure}[t]
\centering
\includegraphics[width=\columnwidth]{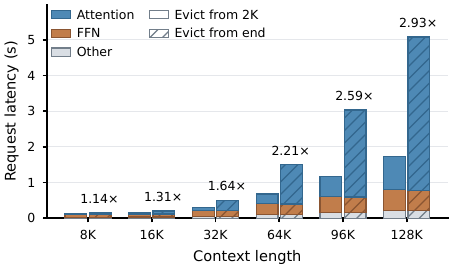}
\caption{Request latency after evicting about 30\% of cached tokens
starting at position 2K or from the end (Qwen3-8B, one H100). Labels give
the speedup of evicting from 2K.}
\label{fig:position-latency}
\end{figure}

Agent workloads amplify the cost asymmetry but also make it harder
to exploit. An LLM agent~\cite{yao2023react,yang2024sweagent} issues
successive model requests while
carrying out a task, incorporating previous interactions and tool
results into later requests, so its history grows and is reused across
turns. Three characteristics of these workloads call for revisiting
position-aware eviction. First, agent workloads involve long contexts.
In the TraceLab
coding-agent dataset~\cite{zhu2026tracelab}, the median request contains
approximately 132K input tokens, and 79\% of requests exceed 64K tokens.
At these lengths, the position of evicted tokens matters more: in a
single-request experiment with Qwen3-8B on one H100 GPU, evicting 30\%
of the cached tokens starting at position 2K rather than from the end
speeds up the next request by 1.14$\times$ at 8K tokens but
2.93$\times$ at 128K (Figure~\ref{fig:position-latency}). Second, agent workloads widely share prefixes.
Requests to the same agent carry identical system prompts and tool
definitions~\cite{srivatsa2025preble,zhang2026cachescout}, and
multi-agent workflows invoke specialized agents that share long fixed
prompts~\cite{pan2025kvflow}. In Claude Code traces from Trace
Commons~\cite{tracecommons2026}, sessions begin with a shared prefix of
about 15.6K tokens, so the beginning of a history is often content that
other sessions reuse. Third, long contexts quickly exhaust GPU memory,
making CPU offloading a standard way to extend cache
capacity~\cite{gao2024cachedattention,qin2025mooncake}. The eviction
policy must account for the two-tier GPU--CPU cache rather than GPU
memory alone.

We therefore build on position-aware eviction with
\emph{head-first eviction}, which retains LRU-based history selection
but evicts KV from the beginning of the selected history rather than its
end. Applying it to agent workloads requires addressing three challenges. First, evicting from the
beginning may remove shared prefixes, reducing prefix reuse and causing
additional recomputation, so the system must locate where the shared
prefix ends and evict only after it.
Second, evicting an earlier segment leaves a
gap before the retained suffix. Conventional prefix-cache lookup
identifies only the contiguous cached prefix, so reusing the suffix
requires the system to track and recompute missing segments without
repeating computation for tokens whose KV cache remains available.
Third, with CPU offloading, independent eviction decisions in GPU and CPU memory can
disrupt the intended head-first eviction order, requiring coordination
across the two tiers.

To address these challenges, we present \sysname, a serving system that
co-designs the eviction policy, KV recomputation, and unified GPU--CPU
cache management around head-first eviction. First, \sysname preserves
shared prefixes and applies head-first eviction to the remaining
history, releasing only the blocks needed. Second, it tracks missing segments and selectively recomputes
them while reusing the retained suffix. Third, it manages GPU and CPU
memory as a single logical cache, selecting segments for eviction across
both tiers and coordinating data migration so that moving cached
segments does not introduce additional evictions.

We implement \sysname in SGLang and evaluate it on three Qwen3 model
sizes using real agent traces under GPU-only and
GPU--CPU configurations. Compared with SGLang, \sysname reduces mean time to first token (TTFT)
by up to 60\% and job completion time (JCT) by up to 37\% with a
GPU-only cache, and by up to 76\% and 60\% with a two-tier GPU--CPU
cache.

This paper makes the following contributions:
\begin{itemize}
  \item Analyze how agent workloads amplify the position asymmetry of
    recomputation cost, and identify the challenges of exploiting it
    (Section~\ref{sec:motivation}).
  \item Design \sysname, which evicts from the beginning of histories
    after shared prefixes, recomputes missing segments while reusing
    retained suffixes, and coordinates eviction and migration across
    GPU and CPU memory (Section~\ref{sec:design}).
  \item Implement \sysname in SGLang and evaluate it with real agent
    traces across three model sizes, with GPU-only and two-tier
    GPU--CPU caches (Section~\ref{sec:eval}).
\end{itemize}

\section{Background and Motivation}
\label{sec:motivation}

\subsection{LLM Inference and KV Cache Management}
\label{sec:bg-attn}

Large language model inference consists of a prefill phase that
processes the input prompt and a decoding phase that generates output
tokens autoregressively~\cite{yu2022orca,agrawal2024sarathi}. At each
transformer layer, attention uses
query, key, and value representations of the tokens. Under causal
attention, each token's query attends to the keys and values of
preceding tokens and itself. During prefill, the model computes and
caches the keys and values of the prompt tokens at each layer. During
decoding, each new token attends to this cached history and adds its
own keys and values to the cache. Beyond reuse within a request,
serving systems retain KV cache after a request completes, allowing
subsequent requests with matching prefixes to reuse previously computed
results. Retaining these cached histories reduces repeated computation
but consumes memory that is also needed by active requests.

When cache capacity is insufficient, serving systems must reclaim
space. Existing mechanisms commonly use LRU to prioritize cached
content for reclamation. They evict KV from the ends of cached sequences,
preserving contiguous prefixes for reuse by subsequent requests with
matching prefixes. CPU offloading extends cache capacity
by retaining KV cache in host memory. In this setting, releasing GPU
memory does not necessarily discard cached content: segments retained
in CPU memory can be transferred back when needed. Recomputation
becomes necessary only when the required KV cache is no longer
available in either tier.

\subsection{The Cost of KV Cache Eviction}
\label{sec:bg-cost}
\label{sec:bg-prefix}

KV cache recomputation cost depends on token position. Restoring evicted
KV cache requires processing the corresponding tokens through the model
again. For a fixed model, projections and feed-forward operations have
approximately position-independent arithmetic costs per token. Under
full causal attention, however, each token attends to all preceding
tokens and itself, so later positions require more attention computation.
We approximate the arithmetic cost of recomputing a token at zero-based
position $i$ as
\[
  c(i) \approx c_{\mathrm{fixed}} + c_{\mathrm{attn}}(i+1),
\]
where $c_{\mathrm{fixed}}$ denotes position-independent computation per
token, including linear projections and feed-forward operations, and
$c_{\mathrm{attn}}$ denotes the main attention computation per query--key
position pair.
Assuming the required preceding KV cache is available, the total
arithmetic cost of recomputing a missing segment
of $b$ tokens starting at position $s$ is
\[
  C(s,b) \approx b\,c_{\mathrm{fixed}}
  + c_{\mathrm{attn}}\left(bs+\frac{b(b+1)}{2}\right).
\]
For equally sized missing segments, position-independent computation is
the same; the difference comes from attention over longer histories at
later positions. Evicting an earlier segment can therefore reclaim the
same cache capacity while requiring less computation to restore it.
This analysis assumes full causal attention in every layer.

To assess the relevance of position-dependent recomputation costs to
real workloads, we analyze the input lengths of Claude and Codex
requests in release v0.0.2 of the TraceLab
dataset~\cite{tracelab2026data,zhu2026tracelab}. The analysis covers 665,186
valid requests across 8,041 sessions, without selecting long sessions
or truncating contexts. The median input length is approximately 132K
tokens, and 79\% of requests exceed 64K tokens, motivating an examination
of recomputation overhead at long context lengths. To quantify how this
overhead changes with context length, we compare request latency after
evicting equally sized earlier and trailing KV cache segments. We run
Qwen3-8B on a modified SGLang implementation using one NVIDIA H100 GPU. After caching the full history, we evict
approximately 30\% of its tokens, either from a contiguous segment
starting at position 2K or from the end. We extend the input history by
512 tokens to simulate new content in a subsequent turn, then measure
the request latency with one output token.
Figure~\ref{fig:position-latency} reports mean latencies over three
runs, split into attention, FFN, and the remaining operations; speedup
is latency after eviction from the end divided by latency after
eviction starting at 2K. Both cases recompute the same number of tokens,
so the difference comes from attention, whose cost grows with the
position of each recomputed token. Evicting the earlier segment yields
lower request latency at every tested context length, with speedup
increasing from 1.14$\times$ at 8K tokens to 2.93$\times$ at 128K.
The cost model above predicts these speedups within 4\%, with
$c_{\mathrm{fixed}}/c_{\mathrm{attn}}$ computed from Qwen3-8B's architecture (Appendix~\ref{sec:cost-model-params}).
These results motivate the head-first eviction strategy introduced in
Section~\ref{sec:intro}: evicting KV cache from earlier positions while
retaining later positions, where recomputation is more expensive.

\begin{figure*}[t]
\centering
\includegraphics[width=\textwidth]{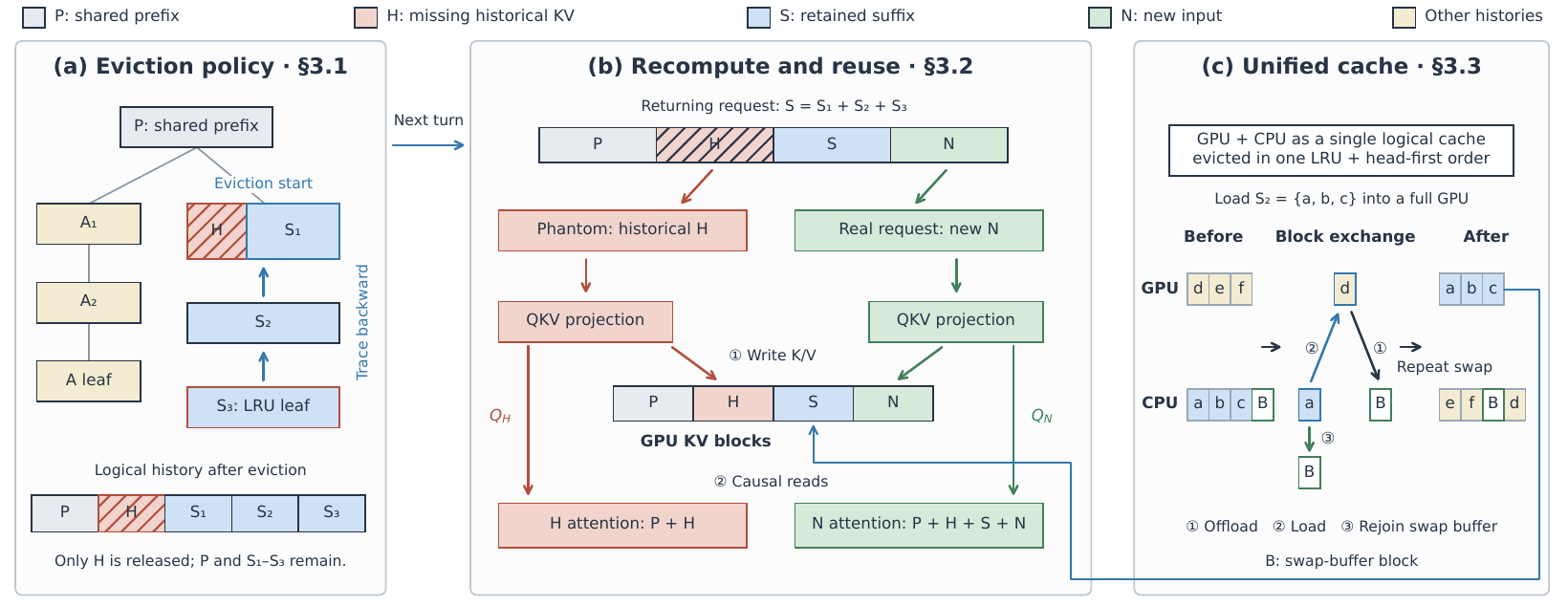}
\caption{\sysname overview. (a) Tracing back from the LRU leaf, eviction
starts after the shared prefix P and releases only the earliest private
segment H. (b) A returning request recomputes H through a phantom request
in the same batch as the real request, reusing the retained suffix S.
(c) GPU and CPU memory form one cache with a single eviction order; when
both tiers are full, blocks are exchanged through a small swap buffer.}
\label{fig:system-overview}
\end{figure*}

\subsection{Challenges}
\label{sec:bg-challenges}

Translating the lower recomputation cost of earlier tokens into serving
performance gains requires addressing three challenges.

First, head-first eviction must preserve shared and in-use KV cache.
The initial prefix may be
shared across requests, so evicting it can remove reusable cache for
other requests and increase recomputation overhead. KV cache needed by
requests currently executing must also remain available. The eviction
boundary therefore depends on both sharing relationships and usage
state: the system must identify the earliest eligible position rather
than simply evicting from the start of the sequence.

Second, head-first eviction leaves a gap in the cached history: an
earlier range of KV cache is missing while a later suffix remains
cached. This affects both cache lookup and request execution.
Conventional prefix matching stops at the gap and cannot locate the
retained suffix. Moreover, conventional execution computes a contiguous
range after the cached prefix. A returning request now has two ranges
to compute: historical tokens whose KV cache is missing and newly
appended tokens. The system must process both ranges efficiently while
reusing the cached suffix between them.

Third, extending head-first eviction to GPU--CPU caches requires
preserving its eviction order across both tiers. Reclaiming space in
each tier involves cache migration or copy removal; selecting candidates
independently within each tier does not ensure that the last copies of
cached KV are removed in the intended order. In write-back caches, for
example, transferring GPU cache to a full CPU cache can trigger CPU
eviction, discarding later-position KV while retaining earlier-position
KV from the same history on the GPU. The system must therefore
coordinate migration and copy removal across both tiers so that final
eviction follows the head-first policy regardless of cache placement.

\section{Design}
\label{sec:design}

Figure~\ref{fig:system-overview} shows the design of \sysname, which
consists of three components: a KV cache eviction policy that preserves
shared prefixes and releases earlier-position KV cache at block
granularity (\S\ref{sec:head-eviction}); a recomputation mechanism that
restores missing historical KV while reusing the retained suffix
(\S\ref{sec:recomputation}); and unified GPU--CPU cache management that
coordinates eviction and migration to preserve the eviction order
across both tiers (\S\ref{sec:multitier}).

\subsection{Eviction Policy}
\label{sec:head-eviction}

\mytitle{Mechanism}
\sysname combines LRU replacement with head-first eviction. LRU chooses
which history to evict, and head-first eviction decides which portion of
it to evict to address the position-dependent recomputation costs
identified in Section~\ref{sec:bg-cost}. When cache
capacity is insufficient, \sysname selects the least recently used
eligible history and releases KV cache starting at its earliest
eligible position. Shared prefixes and KV cache needed by running
requests are protected. Reclamation proceeds in sequence order, moving
to the next eligible history if necessary, until sufficient capacity is
freed. This preserves later-position KV cache whenever only part of a
history needs to be evicted.

\begin{figure}[t]
\centering
\includegraphics[width=\columnwidth]{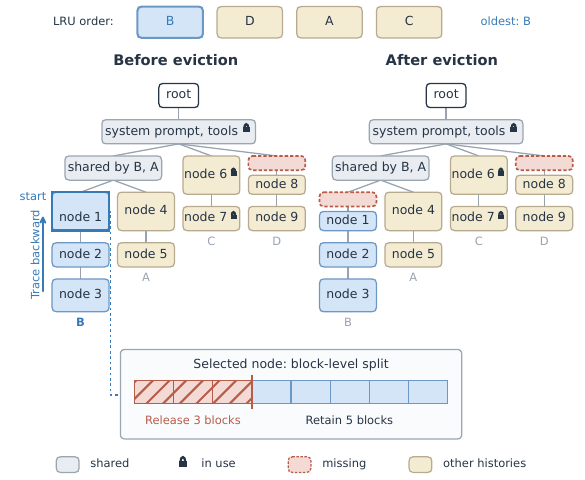}
\caption{Head-first eviction in \sysname. LRU selects history B; tracing
back from its leaf stops below the prefix shared with A, so eviction
starts at node 1, which is split to release only the three blocks
needed.}
\label{fig:eviction-policy}
\end{figure}

To locate the earliest eligible eviction position, \sysname organizes
token histories and their KV cache in a radix tree. Common paths
represent shared prefixes, while each path preserves token order.
Each node stores a contiguous token sequence and its KV cache indices,
together with a reference count tracking use by running requests.
As illustrated in Figure~\ref{fig:eviction-policy}, \sysname traces
backward from the evictable leaf selected by LRU until one of four
conditions holds: \textcircled{\scriptsize 1} the parent is the root,
marking the beginning of the history; \textcircled{\scriptsize 2} the
parent has multiple children, marking a shared-prefix boundary;
\textcircled{\scriptsize 3} the parent has a positive reference count,
indicating that its KV cache is needed by a running request; or
\textcircled{\scriptsize 4} the parent's KV cache is missing, marking
an existing gap. The current node then becomes the starting point for
eviction. Condition \textcircled{\scriptsize 2} also treats branches that
are not shared by active sessions as boundaries, for example a branch
left when a session rewrites its own history, or one shared only with
idle sessions. Such a branch stays protected only until LRU evicts the
stale or idle side, which is then pruned (Section~\ref{sec:implementation}).

\mytitle{Eviction granularity}
After locating the eviction start, \sysname controls how much cache is
reclaimed. SGLang's original reclamation mechanism frees entire radix
nodes, but a node can contain many tokens, particularly near the
beginning of a history where it may hold the initial prompt. Releasing
the entire node can reclaim substantially more cache than needed,
increasing subsequent recomputation. \sysname therefore splits nodes
at boundaries of fixed-size KV cache blocks, releasing only the leading blocks needed
to satisfy the current capacity requirement and retaining the rest; we
call this block-level release, in contrast to SGLang's node-level release.
As illustrated by the inset in Figure~\ref{fig:eviction-policy}, when three blocks
must be reclaimed, \sysname splits the starting node into two parts,
releases its first three blocks, and retains the remaining five blocks.
If the current node cannot satisfy the reclamation requirement,
\sysname continues through subsequent eligible cache in sequence order
until sufficient capacity is freed.

\subsection{Recomputation and Suffix Reuse}
\label{sec:recomputation}
Head-first eviction leaves an earlier range of historical KV cache
missing while retaining the cached suffix. When a subsequent request
extends this history, its input comprises four parts: a cached prefix,
historical tokens whose KV cache is missing, a cached suffix, and newly
appended tokens. Eviction releases KV storage without changing the
historical tokens. With the model and matching token history unchanged,
the retained suffix KV remains valid and does not need to be recomputed.
However, attention for the newly appended tokens still requires the
missing historical KV. \sysname therefore restores only the missing
segment and computes the newly appended tokens, while reusing the cached
prefix and suffix.

To locate the cached suffix despite missing historical KV, \sysname
retains radix-tree nodes' token sequences and parent--child links during
eviction, freeing only the corresponding KV storage and marking it as
missing. Subsequent requests match tokens along the tree path, continuing
through nodes with missing KV as long as their tokens match. Using each
matched node's position and cache state, \sysname identifies the
historical segments to restore and collects the KV indices of reusable
prefixes and suffixes. Missing segments retain only the metadata needed
for matching and are not counted as reusable cache. These segments and
indices are then used to organize recomputation and request execution.

To recompute missing KV while reusing the cached suffix, \sysname
creates an internal prefill-only request, termed a \emph{phantom
request}, for the missing historical segment when assembling the
execution batch. The real request computes
only the newly appended tokens. Both requests preserve their tokens'
original sequence positions and execute in the same batch. \sysname
allocates KV blocks for the missing history and maps both requests to
these blocks. As shown in Figure~\ref{fig:system-overview}(b), at each transformer
layer, \sysname writes the batch's K/V to the cache before computing
attention, so the phantom request writes the missing
K/V before the real request's attention reads them. The real request
can therefore use the restored KV together with the retained prefix
and suffix without waiting for the phantom request to complete a
separate full-model forward pass. The phantom request generates no
output; its restored KV remains in the shared blocks for reuse by the
real request and subsequent requests.

\subsection{Unified GPU--CPU Cache Management}
\label{sec:multitier}

We keep the GPU and CPU tiers exclusive, so that each cached KV segment
resides in only one tier, and manage the two tiers as a single logical
cache.
Within this view, \sysname selects final eviction targets across both
tiers using the LRU and head-first rules in
Section~\ref{sec:head-eviction}, so that physical cache placement does
not alter the eviction order. Migration changes only where cached KV
resides; final eviction occurs only when the last copy of a KV segment
is removed. The system therefore determines which cache to discard
before coordinating migration, rather than selecting eviction targets
independently within each tier.

To reclaim GPU space, \sysname directly frees the selected cache if it
resides on the GPU. If it resides in CPU memory, \sysname evicts it and
offloads retained GPU cache into the freed CPU blocks, thereby reclaiming
GPU space. When a request hits CPU-resident cache that must be loaded
onto the GPU, both the incoming and outgoing cache must be preserved.
If both tiers are full, the GPU has no free blocks for the incoming
cache, and CPU memory has no free blocks for the outgoing cache.
\sysname therefore swaps the two in chunks through a \emph{swap
buffer}, a small pool of reserved CPU blocks.
As illustrated in Figure~\ref{fig:system-overview}(c),
outgoing GPU data is first written into free swap-buffer blocks, and
incoming CPU data is then loaded into the vacated GPU blocks. After the
load completes, the original CPU blocks join the swap buffer for reuse
by subsequent chunks; the outgoing data remains in the CPU blocks that
received it. Each chunk thus reuses the CPU blocks vacated by the
previous one.
Transfer dependencies ensure that a block is reused only
after its previous contents have been transferred, allowing loading
and offloading to proceed without additional eviction.

\section{Implementation}
\label{sec:implementation}

We implement \sysname on top of SGLang~\cite{zheng2024sglang}, an
LLM serving system with radix-tree prefix caching and GPU--CPU
hierarchical caching. Our implementation extends its cache management,
request scheduling, and data transfer modules to support head-first
eviction, KV recomputation, and unified GPU--CPU cache management.
These extensions add approximately 2.5K lines of Python code, including
diagnostic checks and alternative eviction policies used for comparison,
and no new GPU kernels. They run with SGLang's tensor and expert
parallelism, chunked prefill~\cite{agrawal2024sarathi}, and CUDA graphs.

\mytitle{Cache structure maintenance}
When the required reclamation is smaller than the KV cache stored in a
radix-tree node, \sysname splits the node at a block boundary and evicts
only the required portion. The resulting nodes inherit the original
node's access time, preventing structural changes from being treated
as new accesses that alter the LRU order. When eviction reaches the end
of a branch and leaves no reusable cache, \sysname removes missing
nodes that are no longer referenced by requests and continues upward,
pruning ancestors that satisfy the same conditions.

\mytitle{Scheduling adaptation}
First, longest-prefix-match (LPM) scheduling prioritizes requests with
more reusable cache; \sysname counts retained suffixes as reusable and
excludes missing segments. Second, \sysname reserves additional entries
in the request mapping table for phantom requests to record their
tokens' KV cache locations, while keeping the concurrency limit for user
requests unchanged. Third, \sysname includes the phantom requests'
recomputation tokens together with the original request's new tokens in
the batch token budget. If they exceed the chunked-prefill budget,
\sysname admits them together and processes them as a single chunked
prefill, recomputing the missing segments before the new tokens.

\mytitle{Transfer synchronization}
\sysname uses separate CUDA streams for GPU-to-CPU offloading and
CPU-to-GPU loading, with CUDA events enforcing transfer dependencies.
Loading into a GPU block waits until its previous contents have been
offloaded. Likewise, an offload that reuses a CPU block from the swap
buffer waits until the preceding load from that block has completed.
Both offloading and loading are asynchronous. Offloaded cache is
logically treated as CPU-resident as soon as
its transfer is issued, so transfers in flight do not change the unified
eviction order; the transfer dependencies ensure that its CPU blocks are
not overwritten before the data arrives. Computation waits for a load to
complete before accessing the loaded KV cache.

\section{Evaluation}
\label{sec:eval}

\subsection{Experimental Setup}
\label{sec:eval-setup}
\begin{figure*}[t]
\centering
\includegraphics[width=\textwidth,trim=0 2pt 0 0,clip]{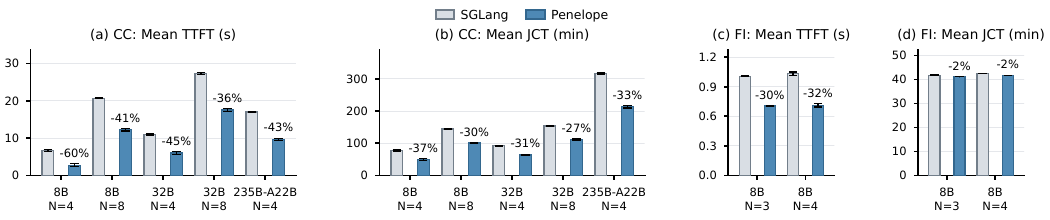}
\caption{Mean TTFT and JCT on agent workloads (mean of three runs; error bars show the range). Bar labels give the change relative to SGLang.}
\label{fig:cc-e2e}
\end{figure*}

\textbf{Testbed.}
We conduct our main experiments on NVIDIA H100 GPUs and evaluate
hardware generalization on NVIDIA A100 GPUs, each with 80 GB of memory.
Each deployment uses one to eight GPUs, depending on the model and
parallelism configuration. Each H100 server has two AMD EPYC 9454
CPUs (48 cores each) and 1.48 TiB of DRAM; each A100 server has two
Intel Xeon Platinum 8380 CPUs (40 cores each) and 0.98 TiB of DRAM. GPUs connect to the host over PCIe Gen4 x16.

\textbf{Models and serving configuration.}
We evaluate three Qwen3 models~\cite{yang2025qwen3}: Qwen3-8B, Qwen3-32B, and
Qwen3-235B-A22B-Instruct-2507-FP8. Unless otherwise specified, we serve
8B on one GPU, 32B with tensor parallelism across four GPUs, and 235B
with tensor and expert parallelism across eight GPUs. We extend the
context length of 8B and 32B to 128K using YaRN~\cite{peng2024yarn} with a scaling factor
of four; the 235B model supports a native context length of 256K.
Prefill uses FlashAttention-3~\cite{shah2024flashattention3}. Decode uses
FlashAttention-3 in GPU-only configurations and
FlashInfer~\cite{ye2025flashinfer} in GPU--CPU configurations, which
SGLang selects automatically when hierarchical caching is enabled; both
systems use the same backends in every comparison.
KV blocks hold 64 tokens. The software stack comprises CUDA 12.8,
PyTorch 2.9.1, sgl-kernel 0.3.21 (which provides FlashAttention-3), and
FlashInfer 0.6.3.
We use $1\mathrm{K}=1024$ tokens. The default GPU-only KV cache
capacities are 256K, 192K, and 384K tokens, respectively. We cap these
capacities to create memory pressure with a manageable number of
concurrent sessions; Section~\ref{sec:eval-sweep} varies the number of
sessions to cover different levels of pressure. Unless otherwise noted, each result is the mean of at
least two runs.

\textbf{Workloads.}
We use two coding-agent trace datasets.
\begin{itemize}
\item \textbf{Claude Code (CC).} Trace Commons~\cite{tracecommons2026}
collects execution traces of Claude Code coding agents, including user
interactions, model replies, and tool calls. We use 24 sessions comprising
4,688 requests. For the initial context missing from the public records,
we reconstruct its length using provider-reported token counts and fill
it with synthetic tokens, identical across sessions for the shared
15.6K-token portion (the median first-turn cache read) and distinct otherwise. After
adding these portions, the median per-turn context length is 80K tokens.
\item \textbf{FreeInference (FI).} This dataset~\cite{freeinference2026trace}
records multi-turn agent requests, including session structure, request
timestamps, and input block identifiers. We use one week of Codex-style
agent traces containing 48 sessions and 4,089 requests, with a median input
length of 77K tokens. During replay, we map each block identifier to a
fixed token sequence to preserve prefix-matching relationships among inputs.
\end{itemize}

The replay maintains $N$ concurrently active sessions. Each session issues
its next request after the previous response completes and the corresponding
wait interval elapses. Agent traces use recorded tool waits. Each session
ends before its first request exceeding the context limit. Subsequent
requests contain the replies recorded in the trace rather than those
generated during replay. All compared systems therefore retain only the
input prompt's KV cache for reuse by subsequent requests and release the
generated tokens' KV cache when the current request finishes.

\textbf{Metrics.}
We report time to first token (TTFT) and job completion time (JCT).
TTFT measures the time from request submission to the first output token.
Each job corresponds to a complete agent session, and JCT measures its
end-to-end duration, including inference, queueing, and tool waits across
turns. We additionally report cache hit rate and tail latency in the
corresponding experiments to examine cache reuse and latency variation
across requests. Hit rate is the fraction of all input tokens served from the KV cache
in either tier.

\textbf{Baselines.}
We compare \sysname with SGLang, an LLM serving system supporting
continuous batching and radix-tree-based prefix caching~\cite{zheng2024sglang}.
We use SGLang at commit \texttt{c36a10aa} with its native LRU eviction
policy and enable CPU offloading for GPU--CPU experiments. All GPU--CPU
experiments use write-back caching. Unless otherwise specified, all
systems use LPM scheduling and are configured
with the same total CPU cache capacity in GPU--CPU experiments.
\sysname reserves 1,024 tokens of CPU cache for its swap buffer, giving
the baseline 1,024 more tokens of usable cache capacity.

\subsection{End-to-End Performance}
\label{sec:eval-e2e}

We evaluate the end-to-end performance of \sysname under GPU-only and
GPU--CPU serving. The GPU-only experiments cover Claude Code
and FreeInference, with additional A100
experiments to assess hardware generalization. The GPU--CPU experiments
further evaluate system performance with caching across GPU and CPU memory.

\mytitle{Agent workloads}
Figure~\ref{fig:cc-e2e} compares \sysname with SGLang on CC across
the 8B, 32B, and 235B models, and on FI with the 8B model. On CC,
\sysname reduces mean TTFT by 36--60\% and mean JCT by 27--37\%.
On FI with three and four concurrent sessions, it reduces mean TTFT by
30\% and 32\% while raising the hit rate by 1.2--2.0 points; JCT
improves by only about 2\% because sessions in this harness spend most of
their time in tool waits, so faster prefill shortens a small part of each
session. These gains combine the effects of eviction granularity and
position: block-level release reduces unnecessary eviction,
while evicting KV cache from earlier positions reduces subsequent
recomputation cost. Section~\ref{sec:eval-ablation} examines their
contributions.

We repeat the 8B CC experiments with four and eight concurrent sessions
on A100-80GB, keeping cache capacity, context limit, replay, and generation
settings the same as in the corresponding H100 configurations.
Table~\ref{tab:hw} shows that \sysname reduces mean TTFT by 52\% and
41\% and mean JCT by 42\% and 35\%, respectively, demonstrating that
\sysname's benefits extend to A100. The JCT reductions are larger than on H100.
Replay records show that the same sessions spend the same absolute time
waiting for tools, but these waits account for only 4--9\% of baseline
JCT on A100, compared with 10--22\% on H100. Consequently, tool waits
dilute the benefit of faster request processing less on A100.

\begin{table}[t]
\centering
\caption{CC on A100-80GB (Qwen3-8B, LPM). Cells report SGLang /
\sysname.}
\label{tab:hw}
\vskip 0.1in
\begin{tabular}{lrr}
\toprule
Concurrency & TTFT (s) & JCT (min) \\
\midrule
$N=4$ & 19.1 / 9.1 & 180 / 104 \\
$N=8$ & 58.1 / 34.1 & 361 / 234 \\
\bottomrule
\end{tabular}
\end{table}

\mytitle{GPU--CPU serving}
\label{sec:eval-cpu}
Table~\ref{tab:gpu-cpu-main} compares \sysname with native SGLang
offloading when the GPU and CPU tiers have equal capacity. SGLang's LPM
scheduler prioritizes requests with the longest cached prefix but counts
only prefixes cached on the GPU, missing histories offloaded to the CPU.
For a fair comparison, both schedulers also count prefixes cached in CPU
memory, which reduces SGLang's mean TTFT by 13--37\%. On top of this
stronger baseline, \sysname further reduces mean TTFT by 46--76\% and
mean JCT by 36--60\%.

\sysname's hit rate exceeds SGLang's by 29--30 points on 8B and 32B and
by 43 points on 235B. The 235B trace also has the longest histories
(144K tokens per request on average, versus 83K for 8B and 32B), so the
tokens SGLang recomputes lie at later positions. Together, these give
\sysname its largest TTFT reduction on 235B, 76\% versus 46--65\% on the
smaller models. With the same trace and tier sizes as 8B and 32B,
235B's hit-rate gap falls to 27 points and its TTFT reduction to 49\%,
within that range.

\begin{table}[t]
\centering
\caption{GPU--CPU serving on CC under LPM, with both schedulers counting
prefixes cached in either tier. Cells report SGLang offloading /
\sysname.}
\label{tab:gpu-cpu-main}
\vskip 0.1in
\begingroup
\setlength{\tabcolsep}{1.5pt}
\begin{tabular}{lcccc}
\toprule
Model & $N$ & \shortstack{GPU / CPU\\capacity (tokens)} & \shortstack{Mean TTFT\\(s)} & \shortstack{Mean JCT\\(min)} \\
\midrule
8B & 4 & 128K / 128K & 10.4 / 4.5 & 81 / 50 \\
8B & 8 & 128K / 128K & 23.5 / 12.8 & 124 / 79 \\
32B & 4 & 128K / 128K & 11.0 / 3.9 & 84 / 47 \\
235B-A22B & 4 & 256K / 256K & 24.2 / 5.8 & 359 / 142 \\
\bottomrule
\end{tabular}
\endgroup
\end{table}

\subsection{Sensitivity Analysis}
\label{sec:eval-sweep}
\begin{figure}[t]
\centering
\includegraphics[width=\columnwidth,trim=0 2pt 0 0,clip]{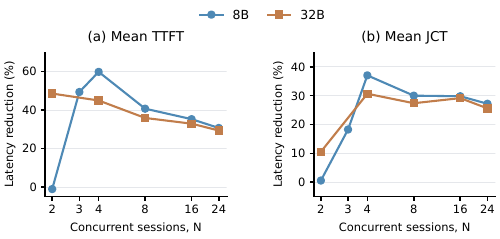}
\caption{Mean TTFT and JCT reductions relative to SGLang as CC concurrency increases.}
\label{fig:concurrency}
\end{figure}

\mytitle{Concurrency and cache pressure} Figure~\ref{fig:concurrency} shows how performance varies with the number
of concurrent CC sessions for the 8B and 32B models. For 8B at $N=2$,
the 256K-token cache accommodates the session histories: both systems
achieve a 99.5\% hit rate, and TTFT and JCT differ by less than 3\%.
Once eviction occurs, \sysname provides substantial benefits, with the
TTFT reduction reaching 60\% at $N=4$. The 32B configuration uses a
smaller 192K-token cache and already experiences eviction at $N=2$,
where TTFT falls by 49\% despite a hit-rate increase of only 2.9 percentage
points. This is consistent with preserving later KV cache that is more
expensive to recompute: improvements in hit rate alone do not fully
capture the recomputation cost saved. As concurrency increases further,
the relative TTFT benefit declines but persists at the highest tested
concurrencies: 31\% for 8B and 29\% for 32B at $N=24$.

JCT improvement also depends on the fraction of session time spent
waiting for the first token. At low concurrency, this fraction is small,
so a large relative TTFT reduction produces a smaller JCT improvement:
for 8B at $N=3$, TTFT falls by 49\% and JCT by 18\%; for 32B at $N=2$,
the reductions are 49\% and 11\%, respectively. As concurrency grows,
first-token waiting accounts for a larger share of session time,
allowing TTFT improvements to translate more fully into JCT savings.
For 8B, the JCT reduction reaches 37\% at $N=4$ and then falls to 27\%
at $N=24$. For 32B, the TTFT reduction declines from 45\% at $N=4$ to 29\% at
$N=24$, while the JCT reduction stays at 26--31\%. This is consistent with
an increasing share of first-token waiting offsetting part of the decline
in relative TTFT benefit.

\mytitle{Scheduler interaction}
\label{sec:eval-sched}
Table~\ref{tab:sched} repeats the $N=4$ CC configuration of each model
under first-come-first-served (FCFS) scheduling.
\sysname reduces mean TTFT and JCT relative to native SGLang under
both schedulers, but the gain is smaller under FCFS: the TTFT reduction
falls from 60\% to 47\% for 8B, from 45\% to 26\% for 32B, and from
43\% to 17\% for 235B. FCFS does not favor sessions whose history is
cached, so a waiting session's history is more often evicted entirely,
and then the eviction position no longer matters. Table~\ref{tab:sched}
shows this directly, and the loss of gain tracks it: under FCFS, the
share of \sysname's requests that recompute their whole private history
rises only from 2\% to 6\% for 8B, but from 9\% to 21\% for 32B and
from 4\% to 21\% for 235B. Native SGLang loses whole histories at similar rates under
both schedulers.

\begin{table}[t]
\centering
\caption{Scheduler comparison on CC at $N=4$. Hit and full-miss rates
are SGLang / \sysname; full miss is the share of requests that recompute
their whole private history. $\Delta$TTFT and $\Delta$JCT are relative to
SGLang under the same scheduler.}
\label{tab:sched}
\vskip 0.1in
\begingroup
\setlength{\tabcolsep}{2pt}
\begin{tabular}{llrrrr}
\toprule
Model & Scheduler & Hit & Full miss & $\Delta$TTFT & $\Delta$JCT \\
 & & (\%) & (\%) & (\%) & (\%) \\
\midrule
8B & LPM & 66.5 / 83.6 & 27.9 / 2.2 & $-60$ & $-37$ \\
 & FCFS & 64.2 / 79.1 & 30.1 / 5.8 & $-47$ & $-31$ \\
32B & LPM & 52.8 / 73.2 & 45.0 / 8.6 & $-45$ & $-31$ \\
 & FCFS & 47.6 / 63.5 & 50.2 / 21.0 & $-26$ & $-19$ \\
235B & LPM & 61.9 / 80.7 & 28.8 / 3.9 & $-43$ & $-33$ \\
 & FCFS & 55.4 / 66.5 & 35.5 / 21.4 & $-17$ & $-13$ \\
\bottomrule
\end{tabular}
\endgroup
\end{table}

\begin{figure}[t]
\centering
\includegraphics[width=\columnwidth]{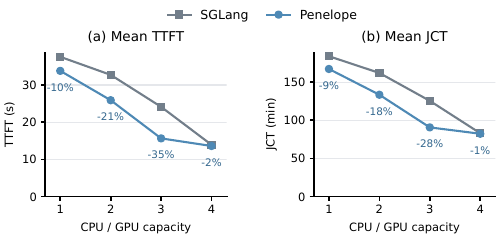}
\caption{Mean TTFT and JCT as the CPU tier grows relative to the GPU
tier (8B, CC, $N=8$, FCFS, 128K-token GPU tier). Labels give the change
relative to SGLang.}
\label{fig:cpu-sweep}
\end{figure}

\mytitle{CPU cache capacity}
Figure~\ref{fig:cpu-sweep} varies the CPU tier from one to four times
the GPU tier, with the 8B model, eight concurrent CC sessions, and a
128K-token GPU tier. Both systems use FCFS so that they process
requests in the same order, isolating the effect of cache management.
The gain of \sysname first grows with the CPU tier and then shrinks,
following the hit rates. At 1:1, the CPU tier is too small and both
systems miss often (hit rates of 42\% for \sysname and 27\% for
SGLang), so TTFT falls by only 10\%. At 3:1, \sysname approaches full
hits (94\%) while SGLang reaches 69\%; the gap is largest, and TTFT and
JCT fall by 35\% and 28\%. At 4:1, SGLang also reaches a 92\% hit rate,
little recomputation remains to be saved, and the two systems have
nearly the same TTFT and JCT.

\subsection{Ablation Study}
\label{sec:eval-ablation}

To separate the contributions of eviction position and granularity and
examine how they work together, we conduct a joint ablation with the 8B
model, four concurrent CC sessions, and LPM scheduling as in the main
comparison. We consider three eviction starts: the tail of the
last node, the head of the last node, and the start of the private
history. Each is paired with either node-level release or block-level
release. The last-node-head configuration tests whether reversing the
direction within the current node suffices, or whether eviction must
trace the cached path back to the private-history start. Last-node-tail
eviction with node-level release corresponds to SGLang; private-history
start with block-level release corresponds to the complete \sysname.
Table~\ref{tab:ablation} reports hit rates and reductions in mean TTFT
and JCT relative to SGLang.

\begin{table}[t]
\centering
\caption{Joint ablation of eviction position and granularity (8B, CC,
LPM, $N=4$). $\Delta$TTFT and $\Delta$JCT are relative to SGLang.}
\label{tab:ablation}
\vskip 0.1in
\begingroup
\setlength{\tabcolsep}{4pt}
\begin{tabular}{lrrr}
\toprule
Eviction start & Hit (\%) & $\Delta$TTFT (\%) & $\Delta$JCT (\%) \\
\midrule
\multicolumn{4}{l}{\textit{Node-level release}} \\
\shortstack[l]{Last-node tail\\(SGLang)} & 64.5 & 0 & 0 \\
Last-node head & 69.0 & $-12$ & $-8$ \\
Private-history start & 66.7 & $-6$ & $-4$ \\
\midrule
\multicolumn{4}{l}{\textit{Block-level release}} \\
Last-node tail & 82.6 & $-45$ & $-28$ \\
Last-node head & 81.5 & $-43$ & $-26$ \\
\shortstack[l]{Private-history start\\(\sysname)} & 84.5 & $\mathbf{-63}$ & $\mathbf{-39}$ \\
\bottomrule
\end{tabular}
\endgroup
\end{table}

With node-level release, changing the eviction start reduces TTFT by at
most 12\%, against 43--63\% with block-level release. Over-released cache must be recomputed on
subsequent requests, increasing space demand and triggering further
evictions, so more requests lose most of their private histories.
Requests that lose at least 90\% of their private history account for
90--94\% of recomputation with node-level release, compared with
16--23\% with block-level release. Once the entire private history is
lost, recovery must recompute it from the beginning regardless of where
eviction started, eliminating the advantage of the eviction position.

Block-level release preserves part of the history, making eviction
position consequential, but only when eviction reaches the start of the
private history. Starting at the last-node tail or the last-node head
reduces TTFT by 45\% and 43\%, a difference within run-to-run variation:
the last node holds the newest tokens of the session, so both ends of it
lie near the end of the context and cost about the same to recompute.
Tracing the cached path back to the private-history start reduces TTFT
by 63\% and JCT by 39\%. Its hit rate of 84.5\% is only 1.9 points above
tail eviction at the same granularity, yet it lowers TTFT by a further
33\% and JCT by 15\%. Thus, the benefit depends on both how much cache is
retained and which positions are retained: block-level release preserves
partial histories, while eviction from the start of the private history
retains the later cache that is most expensive to recompute. The two
mechanisms work together.

\mytitle{Overhead}
We measure the execution overhead of batching a phantom request with
its original request and the CPU overhead of cache management.
With the 8B model on H100 and contexts of 32K--128K tokens, batching
these requests yields an overall execution time within 1\% of computing
the same positions using two separate ordinary requests: attention is
about 1\% slower in the shared batch, while the rest of the forward pass
is slightly faster. Results on A100 are similar.

We also time cache-management operations over one hour of the 8B CC
configuration with $N=8$ and a 256K-token cache, including prefix matching and
identification of missing segments, admission and phantom-request
creation, eviction, and tree insertion. Their cumulative CPU time
accounts for 1.6\% of run time on H100, compared with 0.54\% for
SGLang.

\mytitle{Correctness validation}
Unified cache management should make physical placement irrelevant to
eviction: given the same sequence of requests, \sysname with a
$G$-token GPU tier and a $C$-token CPU tier should hit, retain, and
evict exactly the same KV as with a single $(G{+}C)$-token GPU pool. We
process the requests in both configurations in a fixed order with a
batch size of one, enlarging the CPU tier by the 1,024-token swap buffer
so that its effective capacity is exactly $C$, and compare, after every
request, the cached tokens the request hits, the logical contents
retained across both tiers, and the blocks finally evicted from both
tiers together with their order; moving blocks between tiers is not
counted as eviction.

We run a hand-constructed example that exercises demotion, swap-in, and
eviction across tiers, 20 randomly generated multi-session sequences,
and the CC trace on the 8B model with a 128K-token GPU tier and a
128K-token CPU tier. Every compared quantity matches after every
request. On the CC trace, 602 of the 763 load-backs find both tiers full and exchange blocks through the swap buffer, and the greedily
decoded outputs are identical to those of the GPU-only pool on all
1,245 turns.

We validate the correctness of missing-segment recomputation and suffix
reuse using
the 8B model with greedy decoding. We compare \sysname with a 40K-token
cache against a reference run with a 256K-token cache that never evicts,
with SGLang's deterministic inference mode enabled for both. The test
covers same-batch recomputation, chunked recomputation, and suffix
reuse, at chunk sizes of 8K, 4K, and 2K tokens. In every configuration,
each request's output is bit-identical to the reference, including
the prompt log-probabilities at positions in recomputed missing
segments and reused suffixes.

\section{Related Work}
\label{sec:related}

\mytitle{Prefix caching and tiered KV storage}
vLLM~\cite{kwon2023vllm} manages the KV cache in fixed-size blocks and
reuses blocks keyed by their prefix hash; SGLang~\cite{zheng2024sglang}
indexes cached prefixes in a radix tree.
CachedAttention~\cite{gao2024cachedattention} and
Mooncake~\cite{qin2025mooncake} extend the cache to host memory and
disaggregated storage. Cake~\cite{jin2024cake} computes the start of a
context while loading the rest from slow storage, since later tokens
cost more to compute.

\mytitle{Cost-aware eviction}
Marconi~\cite{pan2025marconi} balances recency against the compute
saved per byte, and RAGCache~\cite{jin2025ragcache} applies a
prefix-aware GreedyDual policy. Pensieve~\cite{yu2025pensieve} evicts
from the leading end of a conversation and recomputes dropped tokens
together with the new prompt. AsymCache~\cite{shi2026asymcache} weighs
hit rate against position-dependent recomputation cost in vLLM, and
concurrent work by \citet{liu2026fancyeviction} favors cheap,
shallow blocks as victims.

\mytitle{Agent-aware KV management}
InferCept~\cite{abhyankar2024infercept} chooses per tool call whether to
preserve, discard, or swap a paused request's KV, and
Continuum~\cite{li2025continuum} pins a session's KV across tool calls.
KVFlow~\cite{pan2025kvflow} uses the agent workflow to guide eviction
and prefetching, ThunderAgent~\cite{kang2026thunderagent} schedules agent
workflows as programs to maximize KV cache hits, and
CacheScout~\cite{zhang2026cachescout} predicts which cached contexts will
be reused.

Like Pensieve and AsymCache, \sysname exploits the position asymmetry
of recomputation cost, but it targets agent workloads: it starts
eviction after prefixes shared across sessions, recomputes missing
segments while reusing retained suffixes, and preserves one eviction
order across exclusive GPU and CPU tiers.

\section{Conclusion}
\label{sec:conclusion}

The position of evicted tokens within a history determines how much
recomputation an eviction costs: in agent workloads, later tokens attend
to longer histories and are more expensive to recompute, yet existing
serving engines evict them first.
We present \sysname, which preserves shared prefixes and evicts from the
beginning of private histories. It combines block-level release,
recomputation of missing segments with suffix reuse, and unified GPU--CPU cache
management.
Compared with SGLang, \sysname lowers mean TTFT by up to 60\% and mean
JCT by up to 37\% with a GPU-only cache, and by up to 76\% and 60\% with
a two-tier GPU--CPU cache.

\bibliography{references}
\bibliographystyle{mlsys2025}

\clearpage
\appendix
\section{Supplementary Details}
\label{sec:eval-additional}

\subsection{Cost Model Parameters}
\label{sec:cost-model-params}
The cost model in Section~\ref{sec:bg-cost} uses the ratio
$c_{\mathrm{fixed}}/c_{\mathrm{attn}}$, which follows from the Qwen3-8B
architecture (hidden size 4{,}096, 32 query heads and 8 KV heads of
dimension 128, and FFN intermediate size 12{,}288). Per token and layer,
the Q, K, V, and O projections and the three FFN matrices take
$4096\times(4096+1024+1024+4096+3\times 12288)=4096\times 47{,}104$
multiply--accumulate operations (MACs). Per query--key pair and layer,
computing the attention score and accumulating the value take
$2\times 32\times 128=8{,}192$ MACs. The ratio is therefore
$4096\times 47{,}104/8{,}192=23{,}552$. With this ratio, the model
predicts speedups of 1.11, 1.29, 1.61, 2.14, 2.55, and 2.87$\times$ for
the six context lengths in Figure~\ref{fig:position-latency}, against
measured speedups of 1.14, 1.31, 1.64, 2.21, 2.59, and 2.93$\times$.

\subsection{Claude Code Main Results}
Table~\ref{tab:e2e} lists the per-configuration results behind
Figure~\ref{fig:cc-e2e}, including p90 TTFT and input throughput.
\begin{table}[h]
\centering
\caption{Claude Code under LPM scheduling, mean of three runs, compared on
the turns recorded in all runs of each pair.
CC$n$ denotes $n$ concurrent sessions. Hit is cache hit rate;
TTFT and JCT are in seconds. In is the total number of input tokens,
including cache hits, divided by the wall-clock time of the replay, in
$10^3$ tokens/s.}
\label{tab:e2e}
\vskip 0.1in
\resizebox{\columnwidth}{!}{%
\begin{tabular}{llrrrrr}
\toprule
Workload & System & Hit\% & TTFT & p90 & JCT & In \\
\midrule
8B CC4 & \sysname & 83.6 & 2.7 & 8.9 & 2927 & 31.0 \\
 & SGLang & 66.5 & 6.7 & 24.7 & 4649 & 20.4 \\
8B CC8 & \sysname & 62.2 & 12.3 & 30.8 & 6047 & 21.8 \\
 & SGLang & 43.4 & 20.7 & 51.4 & 8638 & 15.1 \\
32B CC4 & \sysname & 73.2 & 6.0 & 16.0 & 3840 & 23.5 \\
 & SGLang & 52.8 & 11.0 & 29.8 & 5539 & 16.5 \\
32B CC8 & \sysname & 57.0 & 17.5 & 49.6 & 6684 & 19.6 \\
 & SGLang & 35.4 & 27.4 & 58.3 & 9204 & 13.8 \\
235B CC4 & \sysname & 80.7 & 9.8 & 24.0 & 12805 & 26.7 \\
 & SGLang & 61.9 & 17.0 & 51.9 & 19001 & 17.9 \\
\bottomrule
\end{tabular}}
\end{table}

\subsection{Comparison with Block-Level Tail Eviction}
Table~\ref{tab:e2e-full} compares \sysname with tail eviction under the
same block-level release, which frees the same amount of KV from
the same victims but starts at the tail of the last node (the last-node
tail configuration in Section~\ref{sec:eval-ablation}). On 8B and 32B, the hit
rate of \sysname stays within 3.1 points of tail eviction, yet \sysname
reduces mean TTFT by 12--29\% and JCT by 8--17\% and shortens the
replay by 9--18\%. The gain therefore comes mainly from which positions
are retained rather than how much cache is retained. On 235B, \sysname
also retains more: its hit rate is 5.4 points higher, and it reduces
mean TTFT by 25\% and JCT by 18\%.

\begin{table}[!ht]
\centering
\caption{\sysname versus block-level tail eviction on CC under LPM
scheduling. \sysname rows are the means of three runs, as in
Table~\ref{tab:e2e}; tail-eviction rows are the means of two runs. Hit
is cache hit rate; TTFT and JCT are in seconds. In is defined as in
Table~\ref{tab:e2e}.}
\label{tab:e2e-full}
\vskip 0.1in
\resizebox{\columnwidth}{!}{%
\begin{tabular}{llrrrrr}
\toprule
Workload & System & Hit\% & TTFT & p90 & JCT & In \\
\midrule
8B CC4   & \sysname & 83.6 & 2.7 & 8.9 & 2927 & 31.0 \\
         & Tail     & 83.6 & 3.8 & 11.7 & 3436 & 26.3 \\
8B CC8   & \sysname & 62.2 & 12.3 & 30.8 & 6047 & 21.8 \\
         & Tail     & 59.1 & 16.0 & 37.1 & 7247 & 17.9 \\
32B CC4  & \sysname & 73.2 & 6.0 & 16.0 & 3840 & 23.5 \\
         & Tail     & 72.5 & 7.3 & 18.0 & 4283 & 21.3 \\
32B CC8  & \sysname & 57.0 & 17.5 & 49.6 & 6684 & 19.6 \\
         & Tail     & 57.1 & 19.8 & 51.5 & 7291 & 17.7 \\
235B CC4 & \sysname & 80.7 & 9.8 & 24.0 & 12805 & 26.7 \\
         & Tail     & 75.3 & 13.0 & 29.1 & 15648 & 22.3 \\
\bottomrule
\end{tabular}}
\end{table}

\subsection{Cache Equivalence Details and Negative Control}
\label{sec:correctness-control}
This section supplements the cache-equivalence validation in Section~\ref{sec:eval-ablation}. Cached data are identified by session and token range. The hand-constructed example contains 18 requests; the 20
random sequences use fixed seeds with varying session counts, lengths,
and capacities and contain 699 requests in total; and the CC trace runs
four sessions with 1,245 requests. Their hit rates are 33.67\%, 60.13\%,
and 69.71\%, respectively, identical in both configurations.
As a negative control, we rerun all 22 sequences with the CPU tier
reduced by 256 tokens while keeping the GPU-only capacity unchanged.
All of them diverge, confirming that the comparison detects this
capacity mismatch.

\end{document}